\documentclass[%
superscriptaddress, %
twocolumn, %
showpacs,
 amsmath,amssymb,
 aps,
prl,
floatfix, longbibliography ]{revtex4-1}
\usepackage{graphicx}
\usepackage{dcolumn}
\usepackage{soul,color}

\usepackage[colorlinks,linkcolor=blue,anchorcolor=blue,citecolor=blue,urlcolor=blue]{hyperref}
\usepackage{bm}
\usepackage{amsmath,amsfonts,amsthm} 
\begin{document}

\title{
       \hfill {\normalsize\rm Phys. Rev. B {\bf 98} (2018)}\\
       Is Twisted Bilayer Graphene Stable under Shear?
}

\author{Xianqing Lin}
\affiliation{Physics and Astronomy Department,
             Michigan State University,
             East Lansing, Michigan 48824, USA}
\affiliation{College of Science,
             Zhejiang University of Technology,
             Hangzhou 310023, China}

\author{Dan Liu}
\affiliation{Physics and Astronomy Department,
             Michigan State University,
             East Lansing, Michigan 48824, USA}

\author{David Tom\'anek}
\email[E-mail: ]{tomanek@pa.msu.edu}
\affiliation{Physics and Astronomy Department,
             Michigan State University,
             East Lansing, Michigan 48824, USA}

\date{\today}

\begin{abstract}
In twisted bilayer graphene (TBLG), extremely small deviations
from the magic twist angle $\theta_m{\approx}1.08^\circ$ change
its electronic structure near the Fermi level drastically, causing
a meV-wide flat band to appear or disappear. In view of such
sensitivity to minute structural deformations, we investigate the
combined effect of shear and atomic relaxation on the electronic
structure. Using precise experimental data for monolayer and
bilayer graphene as input in a simplified formalism for the
electronic structure and elastic energy, we find TBLG near
$\theta_m$ to be unstable with respect to global shear by the
angle $\alpha{\approx}0.08^\circ$. In TBLG, the effect of shear on
the electronic structure is as important as that of atomic
relaxation. Under optimum global shear, calculated $\theta_m$ is
reduced by $0.04^\circ$ and agrees with the observed value.
\end{abstract}

%


\maketitle


\section{Introduction}

Theoretically postulated~\cite{%
{Bistritzer12233},%
{Morell2010},%
{LopesdosSantos2012},%
{Fang2016}} drastic changes in the electronic structure at the
Fermi level $E_F$ of bilayer graphene (BLG) near the `magic' twist
angle $\theta_m{\approx}1.08^\circ$ have been recently confirmed
experimentally~\cite{{Cao2018},{cao2018unconventional}} and
ignited a feverish research effort in twisted BLG (TBLG).
Superconductivity~\cite{cao2018unconventional} and strongly
correlated electronic behavior~\cite{Cao2018}, observed at
$\theta_m$, are associated with a meV-wide flat band around the
Dirac point~\cite{Bistritzer12233} at $E_F$ with a vanishing
density of states (DOS). This flat band, which is formed only
within an extremely narrow range $\Delta\theta<0.1^\circ$ around
$\theta_m$, is separated by band gaps above and below from the
rest of the electronic
spectrum~\cite{%
{Bistritzer12233},%
{Cao2016},%
{Cao2018},%
{cao2018unconventional}}. From the viewpoint of atomic structure,
nonzero twist causes a Moir\'{e} pattern with domains of AB, BA
and AA stacking to change rapidly in size especially at small
values of the twist angle $\theta$. In view of the unusual
sensitivity of the electronic structure to twist angle $\theta$
alone, we study here the effect of two other soft deformation
modes, namely global shear and atomic relaxation in TBLG.
Published data suggest that shear does affect the electronic
structure of untwisted and unrelaxed BLG~\cite{San-Jose2012}, but
do not report associated energy changes. Many calculations have
explored atomic relaxations in untwisted~\cite{Popov2011} and
twisted BLG and their effect on the electronic structure~\cite{%
Uchida2014,%
Zhou2015,%
Wijk2015,%
Dai2016,%
Jain2017,%
Nam2017,%
zhang2018structural,%
Gargiulo2018,%
carr2018relaxation}, but ignored inhomogeneities in the stacking
structure observed by high-resolution transmission electron
microscopy (TEM)~\cite{{McEuenBLG13},{yoo18}}.

Here we combine continuum elasticity theory with a tight-binding
description of the electronic structure~\cite{DT268} to study the
behavior of BLG under combined twist and shear. We focus on
geometries with a twist angle $\theta$ near the observed magic
angle $\theta_m{\approx}1.08^\circ$, where the electronic
structure, including the appearance and disappearance of a flat
band near $E_F$ as well as band gaps above and below, shows
extreme sensitivity to structural deformations.
We find TBLG near $\theta_m$ to be energetically unstable with
respect to global shear by the angle $\alpha{\approx}0.08^\circ$.
Also, we find that the effect of shear on the electronic structure
is as important as that of atomic relaxation. Under optimum global
shear, calculated $\theta_m$ is reduced by $0.04^\circ$ and agrees
with the observed value.

\section{Computational Approach}

We use precise experimental data for monolayer and bilayer
graphene as input for our calculations. For unstable geometries,
where such data are not available, we use {\em ab initio} density
functional theory (DFT) as implemented in the \textsc{VASP}
code~\cite{{VASP1},{VASP2},{VASP3}}. We used
projector-augmented-wave (PAW)
pseudopotentials~\cite{{PAW1},{PAW2}} and the {\em{SCAN+rVV10}}
exchange-correlation functional, which provides a proper
description of the van der Waals
interaction~\cite{Klime2011,Peng2016}. We used $800$~eV as the
electronic kinetic energy cutoff for the plane-wave basis and a
total energy difference between subsequent self-consistency
iterations below $10^{-5}$~eV/atom as the criterion for
self-consistency. We reached full convergence, since our
calculations were limited to very small unit cells of AA- and
AB-stacked BLG.

\section{Results}

\subsection{Deformation Modes in Bilayer Graphene}

The softest deformation modes of BLG are relative translation and
rotation of the constituent graphene layers. BLG subject to
uniform twist creates a Moir\'{e} lattice of equilateral
triangles, and rigid displacement only translates this lattice
with no effect on the atomic and electronic structure. Atomic
relaxations including local bending in TBLG, as well as global
shear and stretch of the monolayers, require more energy. Global
stretch, which involves bond stretching, is energetically harder
than global shear, which involves only bond bending. Global shear
creates a stripe pattern of domains~\cite{San-Jose2012}. The
combination of twist and shear has not been explored so far.

\begin{figure*}[t]
\includegraphics[width=1.8\columnwidth]{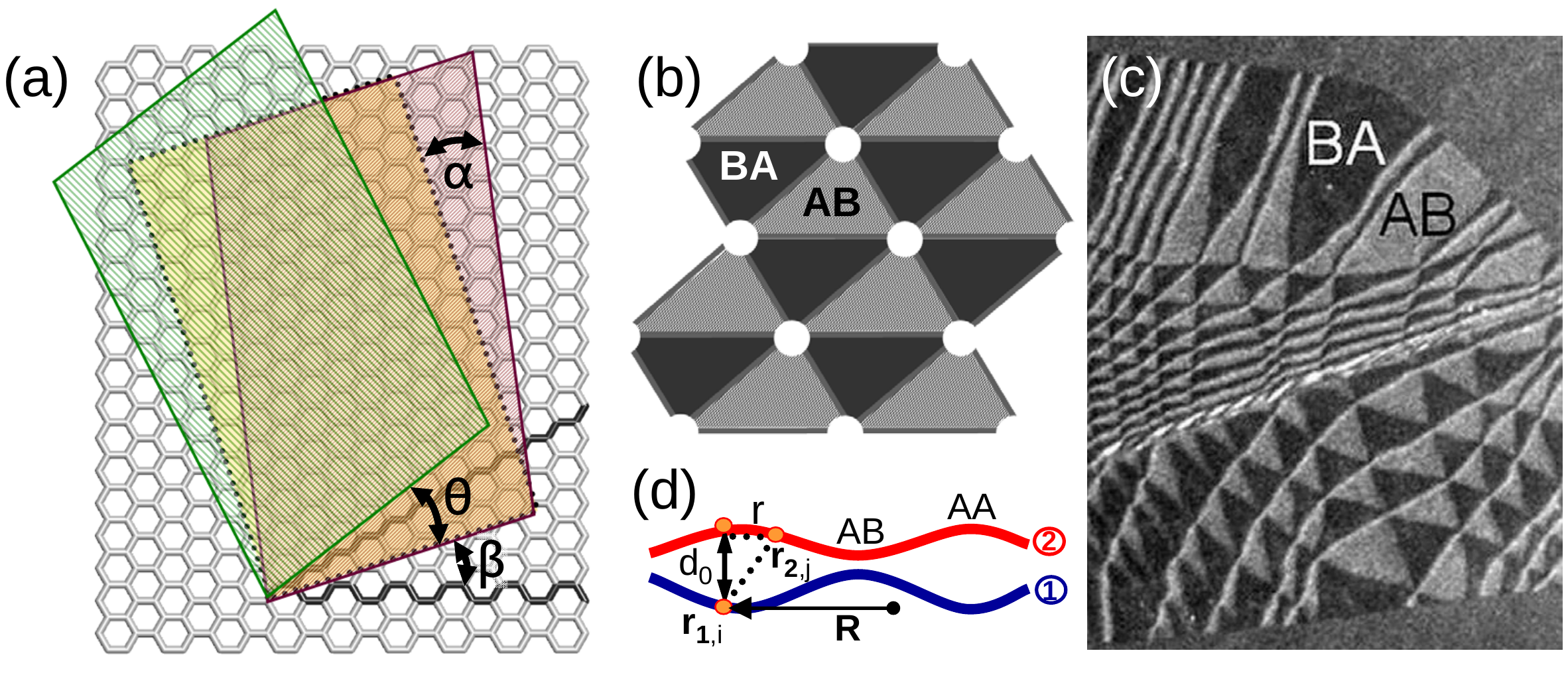}
\caption{Shear and twist in bilayer graphene (BLG). (a) Definition
of shear and twist operations in BLG, initially a bilayer in AA
stacking indicated by the honeycomb lattice. A rectangular segment
of the top layer, with one side closing the angle $\beta$ with
respect to the highlighted armchair direction, is indicated in
yellow and surrounded by the dotted line. The segment is first
sheared by the angle $\alpha$ and subsequently rotated by the
angle $\theta$. %
(b) Schematic top view of a uniformly twisted and sheared
monolayer graphene (MLG) on top of an undeformed MLG. The
resulting relaxed Moir\'{e} pattern contains regions of AA stacking,
highlighted by the white circles, and those of AB or BA stacking. %
(c) Dark-field TEM image of bilayer graphene reproduced from
Ref.~[\onlinecite{McEuenBLG13}]. %
(d) Schematic side view of a relaxed sheared and twisted BLG
with locally varying stacking. %
\label{fig1}}
\end{figure*}

\begin{figure}[b]
\includegraphics[width=1.0\columnwidth]{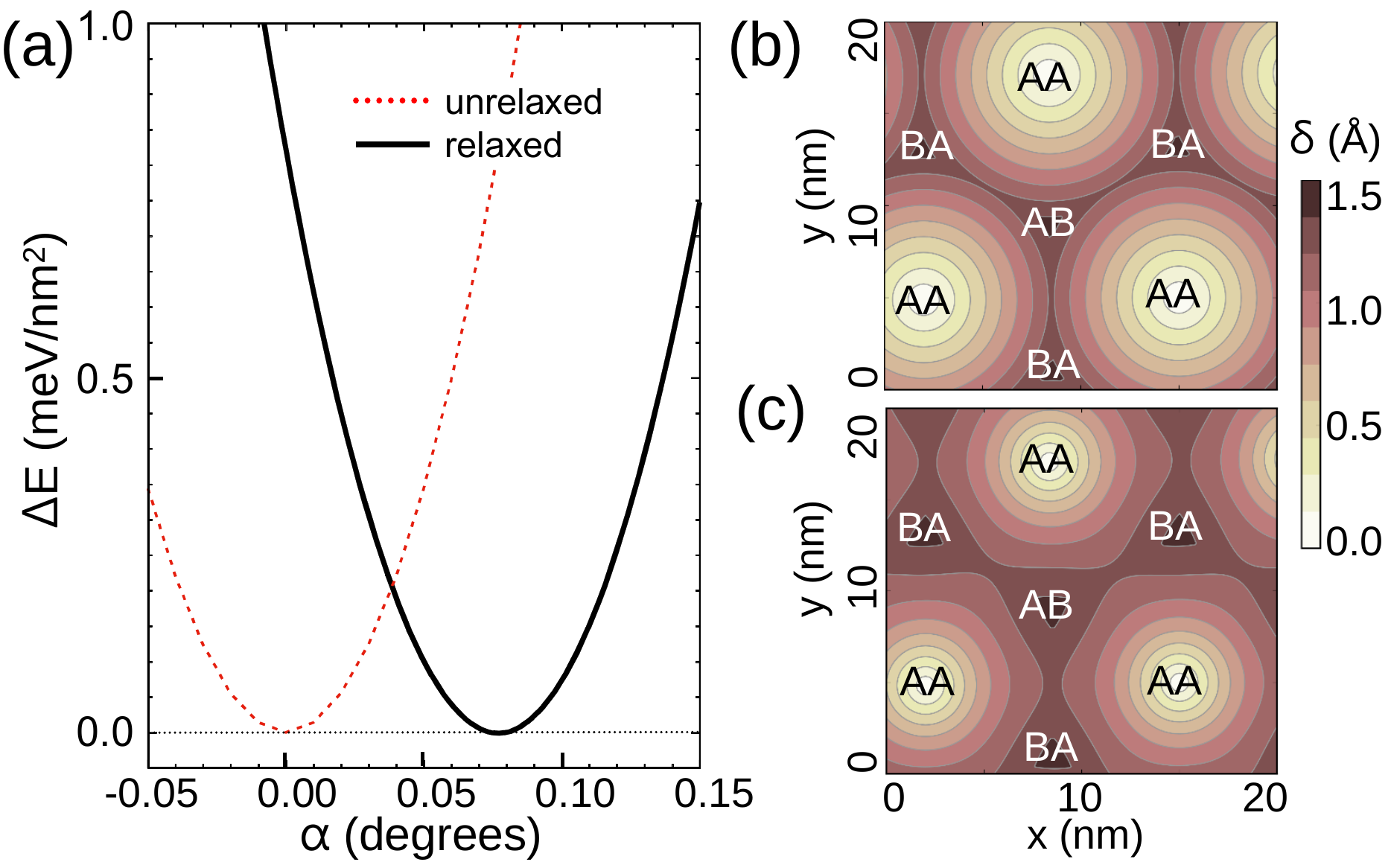}
\caption{%
(a) Energy difference ${\Delta}E$ caused by shearing the top layer
of BLG, which had been twisted by
$\theta_m=1.08^\circ$, %
by varying the shear angle $\alpha$ along the $\beta=0^\circ$
direction. %
Results for the BLG with unrelaxed atomic positions in planar,
sheared monolayers are shown by the %
red dotted line and for the BLG with relaxed atomic positions by
the black solid line. %
Contour plots of the local shift vector length
${\delta}=|\bm{\delta}|$ in the BLG structure, which had been
twisted by $\theta_m=1.08^\circ$ and sheared by
$\alpha=0.08^\circ$ along the $\beta=0^\circ$ direction, %
(b) in absence and %
(c) in presence of lattice relaxation. %
\label{fig2}}
\end{figure}

The deformation of BLG under shear and twist is defined in
Fig.~\ref{fig1}(a). In the initially AA-stacked BLG, we define the
shear direction within the top layer by the angle $\beta$ with
respect to the armchair direction. After first being sheared by
the angle $\alpha$, the to player is subsequently rotated by the
angle $\theta$. The mathematical formulation of the shear-twist
deformation in BLG is detailed in the Appendix.
The general Moir\'{e} pattern of a sheared and twisted BLG is a
lattice asymmetric triangles with AA-stacked regions forming the
vertices, as shown schematically in Fig.~\ref{fig1}(b). The
triangular regions of AB and BA stacking between the vertices have
no symmetry in general, but become equilateral in the case of pure
twist in the BLG associated with ${\alpha}=0^\circ$.
Figure~\ref{fig1}(c) contains a reproduction of a dark-field TEM
image of the BLG reported in Ref.~[\onlinecite{McEuenBLG13}]. The
triangular lattice in this figure is strongly distorted due to
inhomogeneous twist and strain. Quantities associated with atomic
relaxation are defined in the schematic side view of sheared and
twisted BLG in Fig.~\ref{fig1}(d). The local inter-layer distance
$d_0$ depends on the 2D position vector ${\bf{R}}$ within the BLG
plane, which distinguishes regions with different local stacking
sequences.

\subsection{Relaxations in Sheared and Twisted Bilayer Graphene}

To quantify the effect of relaxation in the BLG, we consider both
in-plane and out-of-plane deformations and calculate changes in
the total energy $E_{tot}=E_{el}+E_{int}$ in terms of the elastic
deformation energy $E_{el}$ of the individual layers and the
interlayer interaction energy $E_{int}$. The displacement of the
atom $i$ at ${\bf{r}}_{n,i}$, defined in Fig.~\ref{fig1}(d) for
layers $n=1,2$, is described by a continuous displacement field
with an in-plane component ${\bf{u}}^{(n)}(\bf{R})$ and an
out-of-plane component $h^{(n)}({\bf{R}})$. The elastic
deformation energy is given by~\cite{Andres2012}
\begin{eqnarray}
&E_{el}& = \! \sum_{n=1}^{2} \int d{\bf{R}} \biggl\{
           \frac{\kappa}{2}
           \left(\frac{\partial^2h^{(n)}}{\partial x^2}
            + \frac{\partial^2h^{(n)}}{\partial y^2}\right)^2
            \nonumber \\
&+&\!\frac{\lambda+\mu}{2}
            \left(\frac{\partial u_x^{(n)}}{\partial x} \!+\!
            \frac{\partial u_y^{(n)}}{\partial y}\right)^2
            \nonumber \\
&+&\!\frac{\mu}{2} \left[
         \left(\frac{\partial u_x^{(n)}}{\partial x} -
          \frac{\partial u_y^{(n)}}{\partial y}\right)^2 %
          \!\!\! + \!
         \left(\frac{\partial u_y^{(n)}}{\partial x} +
          \frac{\partial u_x^{(n)}}{\partial y}\right)^2
          \right] \biggr\} ,%
\label{eq1}
\end{eqnarray}
where the integral extends over the Moir\'{e} pattern supercell.
We use $\kappa=1.4$~eV for the flexural rigidity\cite{DT255}, and
$\lambda=4.23$~eV/{\AA}$^2$ and $\mu=9.04$~eV/{\AA}$^2$ for the 2D
elastic Lam\'{e} factors~\cite{carr2018relaxation} of graphene.

To calculate the interlayer interaction energy $E_{int}$, we first
characterize the local stacking at position
${\bf{R}}={\bf{r}}_{1,i}$ of atom $i$ in layer 1, shown in
Fig.~\ref{fig1}(d), by calculating the connection vectors
${\bf{r}}_{2,j}-{\bf{r}}_{1,i}$ to all atoms $j$ of the same
sublattice in layer 2 and their projections onto the $x-y$ plane
of the lattice. The shortest among these vectors is then called
the shift vector~\cite{Popov2011} ${\bm{\delta}}({\bf{R}})$. With
this definition, $|{\bm{\delta}}|=0$ in AA-stacked and
$|{\bm{\delta}}|=a/\sqrt{3}$ in AB-stacked BLG, where
$a=2.46$~{\AA} is the lattice constant of graphene. The observed
interlayer distance in the stable AB-stacked BLG is
$d_0^{AB}=3.35$~{\AA}. The corresponding calculated value for
AA-stacked BLG, which is less stable by
${\Delta}E=0.38$~eV/nm$^2$, is $d_0^{AA}=3.60$~{\AA}.

The interlayer interaction energy depends on the local interlayer
separation $d_0({\bf{R}})$ and the local shift vector
${\bm{\delta}}({\bf{R}})$. Since the Moir\'{e} supercells are much
larger than the interatomic distance in the twist angle range of
interest, both quantities can be represented well by a continuous
field. Then, the interlayer interaction energy is given by the
integral
\begin{equation}
E_{int} = \int V({\bf{R}}) d{\bf{R}}\,, %
\label{eq2}
\end{equation}
which extends over the Moir\'{e} supercell.
We represent the interlayer interaction potential
$V({\bf{R}})=V[{\bm{\delta}}({\bf{R}}),d_0({\bf{R}})]$, as well as
the quantities $d_0$ and $\bm{\delta}$, by a Fourier expansion
over the BLG lattice~\cite{Zhou2015}, which is detailed in the
Appendix.
The expansion requires only few of the shortest reciprocal
superlattice vectors of the BLG and is trivial for an untwisted
BLG with AA and AB stacking, where $V=\textrm{const.}$ and
$V^{AA}-V^{AB}=0.38$~eV/nm$^2$.

The optimum geometry of the relaxed BLG that had been subject to
global shear and twist is obtained by globally minimizing the
total energy, which has been determined using a Fourier expansion
detailed in the Appendix.

Whereas lattice relaxation is driven by energy gain, shear
distortion of a monolayer requires energy investment. For a given
twist angle $\theta$, it is conceivable that the energy invested
in shear may be outweighed by an additional energy gain associated
with relaxation to a more favorable structure. This situation is
illustrated in Fig.~\ref{fig2}(a) that shows energy changes in
TBLG with $\theta_m=1.08^\circ$. We found the energy values, which
are shown for shear along the ${\beta}=0^\circ$ direction, to be
nearly identical for shear along ${\beta}=30^\circ$, an conclude
that the shear energy does not depend on ${\beta}$. When lattice
relaxation is suppressed, as shown by the dotted line, the most
stable geometry is unsheared with ${\alpha}=0^\circ$. When
allowing for lattice relaxation,
the optimum geometry has acquired a global shear angle
${\alpha}=0.08^\circ$.

The absolute value of the local shift vector
${\delta}=|\bm{\delta}|$ as a function of ${\bf{R}}=(x,y)$ is
shown in Figs.~\ref{fig2}(b) and \ref{fig2}(c) for BLG structures
that have been twisted by $\theta_m=1.08^\circ$ and sheared by the
small angle $\alpha=0.08^\circ$ along the $\beta=0^\circ$
direction. The corners of the Moir\'{e} supercell, shown in white,
are the unshifted AA regions with ${\delta}=0$. In the unrelaxed
structure of Fig.~\ref{fig2}(b), the energetically favorable
regions of AB or BA stacking are rather small. Upon relaxation,
these favorable stacking regions increase in size, as seen in
Fig.~\ref{fig2}(c). The effect of relaxation becomes much more
visible at smaller twist angles $\theta$ associated with very
large Moir\'{e} domains. As seen in Fig.~\ref{fig4},
the AB and BA domains then acquire a distinctly triangular shape
upon relaxation, which has been observed in TEM
images~\cite{{McEuenBLG13},{yoo18}} including Fig.~\ref{fig1}(c).

\subsection{Electronic Structure of Sheared,
            Twisted and Relaxed Bilayer Graphene}

To study the combined effect of shear, twist and atomic relaxation
on the electronic structure of BLG, we use an extension of the
minimum Hamiltonian~\cite{DT268} that had successfully reproduced
the electronic structure of twisted BLG. Due to the high in-plane
stiffness and flexural rigidity of graphene, the atomic relaxation
is rather small and smooth across the BLG lattice, so that the
intra-layer nearest-neighbor Hamiltonian with
$V^0_{pp\pi}=3.09$~eV is not affected. Since the interlayer
separation changes between the value $d_0^{AA}$ and $d_0^{AB}$
within each Moir\'{e} domain, as seen in Fig.~\ref{fig1}(d), we
modify the expressions for the interlayer hopping in
Ref.~[\onlinecite{DT268}] to
%
\begin{equation}
t(r,{\bf{R}}) = V^0_{pp\sigma}(d_0)
      e^{-(\sqrt{r^2+d_0^2}-d_0)/\lambda} %
      \frac{d_0^2}{r^2+d_0^2} \,,%
\label{eq3} %
\end{equation}
where
\begin{equation}
V^0_{pp\sigma}(d_0) = V^0_{pp\sigma}(d_0^{AB})
     e^{-(d_0-d_0^{AB})/\lambda'} \,.
\label{eq4}
\end{equation}
In these expressions, the only quantity that depends on the
position ${\bf{R}}$ within the layer is the local interlayer
separation $d_0=d_0({\bf{R}})$, defined in Fig.~\ref{fig1}(d). For
the unrelaxed geometry with $d_0=d_0^{AB}={\rm const.}$, the
values $V^0_{pp\sigma}(d_0^{AB})=0.39$~eV, $d_0^{AB}=3.35$~{\AA},
$\lambda=0.27$~{\AA} have been established in
Ref.~[\onlinecite{DT268}]. We furthermore use the parameter
$\lambda'=0.58$~{\AA} to adequately describe the dependence of
$V^0_{pp\sigma}$ on the interlayer separation to match our DFT
calculations.

In TBLG, where atomic relaxation has not been considered
explicitly in a related previous study~\cite{DT268}, we determined
the electronic structure at the interlayer distance $d_0^{AB}$
using the same continuum method for the description of
eigenstates. %
In unrelaxed TBLG subject to shear, %
the interlayer Hamilton matrix elements of
Ref.~[\onlinecite{DT268}] are modified as
%
%
%
\begin{eqnarray}
&&\langle
\psi_{1,\eta}({\bf{k_1}})|H|\psi_{2,\xi}({\bf{k_2}})
\rangle =%
\nonumber \\
&&\sum_{{\bf{G}},{\bf{G}'}} %
  \frac{\tilde{t}(|{\bf{k_1}}+{\bf{G}}|)}{\Omega} %
  e^{i({\bf{G}}{\cdot}{\bf{\tau}_{\eta}}-{\bf{G}'}{\cdot}{\bf{\tau}_{\xi}'})}
  \delta_{{\bf{k_2}}-{\bf{k_1}},{\bf{G}}-{\bf{G'}}} \,. %
\label{eq5}
\end{eqnarray}
Here we use $\eta$ and $\xi$ to represent the two sublattices with
the basis vectors ${\bm{\tau}}$ in the unsheared bottom layer~1
and ${\bm{\tau}'}$ in the top layer~2 that has been sheared and
twisted. ${\bf{k_n}}$ is the momentum vector in layer~$n$.
$\tilde{t}(k)$ is the 2D Fourier transform of $t(r)$ that is
independent of
the position within the unrelaxed bilayer, which is kept at the
constant optimum interlayer separation $d_0^{AB}$.
${\bf{G}}$ is the reciprocal lattice vector of the bottom layer~1
and ${\bf{G}'}$ is the reciprocal lattice vector of the sheared
and twisted top layer~2. For non-specific values of $\alpha$,
$\beta$ and $\theta$, the BLG lattice is generally incommensurate,
but can be approximated by a commensurate Moir\'{e} superlattice
with large unit cells.

For small twist angles, the reciprocal lattice vectors of the
undeformed bottom layer~1 in Eq.~(\ref{eq5}) can be approximated
by ${\bf{G}}=n_1{\bf{b}}_1+n_2{\bf{b}}_2$ and those of the
deformed top layer~2 by ${\bf{G}'}=n_1{\bf{b}}_1'+n_2{\bf{b}}_2'$,
with $n_1$ and $n_2$ being small integers. In these expressions,
${\bf{b}_{1/2}}$ are the two vectors spanning the reciprocal
lattice of the bottom layer and ${\bf{b'}_{1/2}}$ those spanning
the reciprocal lattice of the deformed top layer.
In relaxed commensurate BLG structures, %
we do not use Eq.~(\ref{eq5}), but rather diagonalize the
tight-binding Hamiltonian directly %
to obtain the electronic band structure.

\begin{figure}[t]
\includegraphics[width=0.9\columnwidth]{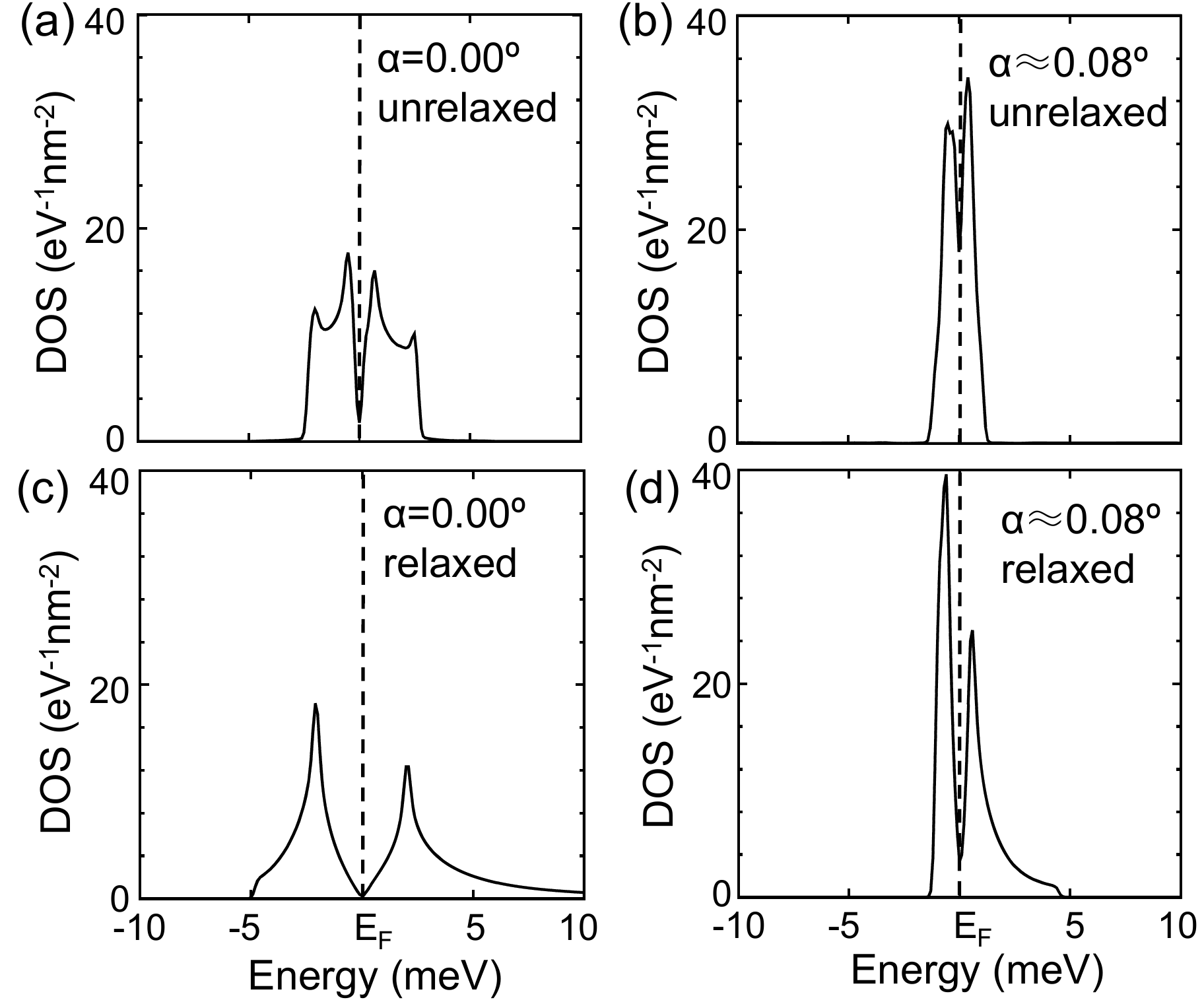}
\caption{Electronic density of states (DOS) in the BLG structure
subject to the magic twist angle $\theta_m=1.08^\circ$. Results
for the unsheared structure are shown in panels (a) and (c), and
those for the top layer sheared by $\alpha=0.08^\circ$ along the
$\beta=0^\circ$ direction in (b) and (d). Results for the
unrelaxed structure in (a) and (b) are compared to those for the
relaxed structure in (c) and (d). %
\label{fig3}}
\end{figure}

The effect of shear and atomic relaxation on the DOS of TBLG
subject to twist by the magic angle $\theta_m=1.08^\circ$ is
discussed in Fig.~\ref{fig3} for energies close to $E_F$ and in
Fig.~\ref{fig5}
for a wider energy range. The DOS characteristics at $\theta_m$
are a narrow `flat band' around the Dirac point with vanishing DOS
at $E_F$ and band gaps above and below.
The DOS of unsheared and unrelaxed TBLG in Fig.~\ref{fig3}(a) is
the same as in a previous report~\cite{DT268} for the same
structure, where shear and atomic relaxation have not been
considered explicitly. Subjecting BLG to global shear by
$\alpha=0.08^\circ$, while suppressing any atomic relaxation,
reduces the width of the flat significantly, as seen in
Fig.~\ref{fig3}(b), to almost half its value in unsheared TBLG.
This result alone proves that even minor shear plays a significant
role in the electronic structure of TBLG. As seen in
Fig.~\ref{fig3}(c), atomic relaxation alone increases the width of
the band near $E_F$ significantly with respect to the structure in
Fig.~\ref{fig3}(a) to the degree that the designation `flat band'
is no longer appropriate. Subjecting this relaxed structure to
minor shear of $\alpha=0.08^\circ$, however, narrows down this
band to resemble that of Fig.~\ref{fig3}(a) for TBLG irrespective
of shear and relaxation. Thus, taking shear into account is
essential to properly identify the $\theta_m$ value in relaxed
TBLG. We compared the width of the narrow band around $E_F$ for
unsheared and sheared TBLG with atomic relaxation. Our results,
reproduced in Fig.~\ref{fig6},
suggest that the minimum bandwidth is found at the magic angle
$\theta_m=1.12^\circ$ in relaxed TBLG with suppressed shear. This
value is larger than the magic angle value $\theta_m=1.08^\circ$
that is observed in sheared and relaxed TBLG.

\section{Discussion}

As reported in the discussion of shear in Fig.~\ref{fig2}(a), the
deformation energy ${\Delta}E$ is nearly independent of the shear
direction $\beta$. Nevertheless, $\beta$ modifies strongly the
shape of the triangles in the Moir\'{e} superlattice. Local
changes in the shear angle $\alpha$ and direction $\beta$ are
clearly visible in the TEM images of BLG reported in
Fig.~\ref{fig1}(c) and Refs.~[\onlinecite{McEuenBLG13}] and
[\onlinecite{yoo18}]. We also note that the high energy cost of
in-layer deformations including shear limits the degree of atomic
relaxation. Distributing shear from one to both layers of BLG
should cut this energy cost in half in the harmonic regime, thus
reducing the limits imposed on atomic relaxation and providing
extra energy gain for the system.

\section{Summary and Conclusions}

In conclusion, we studied the effect of combined shear and twist
on the energy as well as the atomic and electronic structure of
BLG. We found that the observed drastic changes in the electronic
structure near the Fermi level, caused by minute changes of the
twist angle away from the magic value
$\theta_m{\approx}1.08^\circ$, are strongly affected by lattice
relaxation and global shear. Using precise experimental data for
monolayer and bilayer graphene as input in a simplified formalism
for the electronic structure and elastic energy, we found TBLG
near $\theta_m$ to be unstable with respect to global shear by the
angle $\alpha{\approx}0.08^\circ$. We also found that shear and
atomic relaxation modified the electronic structure of TBLG to a
similar degree. At optimum global shear, the calculated value of
the magic angle $\theta_m$ in relaxed TBLG is reduced by
$0.04^\circ$ to agree with the observed value of $1.08^\circ$.

\section{Appendix}
\setcounter{equation}{0}
\renewcommand{\theequation}{A\arabic{equation}}

\subsection{Mathematical formulation of the rigid shear-twist
             deformation in bilayer graphene}

Before describing the shear-twist deformation in bilayer graphene
(BLG), we need to recall that the honeycomb lattice of graphene
consists of two triangular sub-lattices A and B. The Bravais
lattice of the bottom layer~1 is spanned by the vectors
${\bf{{a}_1}}=a(\sqrt{3}/2,-1/2)$ and
${\bf{{a}_2}}=a(\sqrt{3}/2,1/2)$ in Cartesian coordinates, where
$a=2.46$~{\AA} is the lattice constant of graphene. The basis
vectors of the sublattices are ${\bm{{\tau}_A}}=0$ and
${\bm{{\tau}_B}}=({\bf{{a}_1}}+{\bf{{a}_2}})/3$. Initially, the
top layer~2 of the BLG is on top of layer~1 in AA stacking.

The sheer-twist transformation of the BLG top layer with respect
to the bottom layer is shown schematically in Fig.~\ref{fig1}(a).
Pure shear in the graphene top layer by the angle $\alpha$ along
the shear direction angle $\beta$ with respect to the armchair
direction is described by the transformation matrix
\begin{eqnarray}
S_{\alpha} = \left(
   {\begin{matrix}
   {1-\sin\beta \cos\beta \tan\alpha} & { (\cos\beta)^2 \tan\alpha} \cr
   {-(\sin\beta)^2 \tan\alpha} & {1+\sin\beta \cos\beta \tan\alpha} \cr
   \end{matrix}}
   \right).\ \ \
\label{eqS1}
\end{eqnarray}
Pure counter-clockwise twist of the top layer by the angle
$\theta$ with respect to the bottom layer is described by the
unitary transformation matrix
\begin{eqnarray}
T_{\theta} = \left(
   {\begin{matrix}
   {\cos\theta} & {-\sin\theta} \cr
   {\sin\theta} &  {\cos\theta} \cr
   \end{matrix}}
   \right) \,.
\label{eqS2}
\end{eqnarray}
The shear-twist operation transforms the two Bravais lattice
vectors ${\bf{{a}_j}}$ of the top layer to
${\bf{{a}_j'}}=T_{\theta}S_{\alpha}{\bf{{a}_j}}$ with $j=1,2$.
Except for specific values of $\theta$, $\alpha$ and $\beta$, the
shear-twisted BLG (STBLG) lattice is incommensurate. For small
values of $\theta$ and $\alpha$, such an incommensurate structure
can be approximated by a commensurate superlattice with with large
Moir\'{e} supercells. The electronic structure of this system can
be obtained to a good accuracy using the continuum method
described in Ref.~[\onlinecite{DT268}].

The reciprocal Moir\'{e} superlattice is spanned by the two vectors %
${\bf{b_1^{(s)}}}={\bf{b}_2}-{\bf{b'}_2}$ and %
${\bf{b_2^{(s)}}}=({\bf{b'}_1}+{\bf{b'}_2})-({\bf{b}_1}+{\bf{b}_2})$.
In our notation, ${\bf{b}_j}$ with $j=1,2$ are the two vectors
spanning the reciprocal lattice of the bottom layer and the primed
vectors ${\bf{b'}_j}$ span the reciprocal lattice of the top
layer. The two reciprocal lattice vectors ${\bf{b_1^{(s)}}}$ and
${\bf{b_2^{(s)}}}$ define the lattice vectors ${\bf{a_1^{(s)}}}$
and ${\bf{a_2^{(s)}}}$ of the direct Bravais STBLG superlattice.
When ${\bf{a_1^{(s)}}}$ and ${\bf{a_2^{(s)}}}$ both belong to the
Bravais lattice of the individual layers, STBLG becomes
commensurate. This condition is fulfilled when %
${\bf{a_1^{(s)}}}=M_1{\bf{{a}_1}}+N_1{\bf{{a}_2}}$ and %
${\bf{a_2^{(s)}}}=M_2{\bf{{a}_1}}+N_2{\bf{{a}_2}}$ with integer
values of $M_1$, $N_1$, $M_2$, and $N_2$. %

The supercell area of the sheared top layer remains the same as
that of the unsheared bottom layer if $M_2+N_2=N_1-1$. The
quantities $M_1$, $N_1$, $M_2$, and $N_2$ define uniquely the
twist angle $\theta$, the shear angle $\alpha$ and the shear
direction angle $\beta$ of the top layer. One geometry near the
first magic angle, characterized by $(M_1,N_1)=(30,31)$ and
$(M_2,N_2)=(-31,61)$, describes an unsheared twisted BLG layer
with $\theta=1.0845^\circ$ and ${\alpha}=0^\circ$. Another similar
geometry, characterized by $(M_1,N_1)=(31,32)$ and
$(M_2,N_2)=(-36,67)$, describes a sheared twisted BLG layer with
$\theta=1.0845^\circ$, $\alpha=0.0872^\circ$, and
$\beta=-0.5423^\circ$. We determine the lattice relaxation and its
effect on the electronic structure of the sheared and twisted BLG
using the commensurate structures.

\subsection{Mathematical background of the relaxation treatment
             in sheared and twisted bilayer graphene}

BLG subject to rigid shear and twist, which is characterized by
$\theta$, $\alpha$ and $\beta$, is further stabilized by atomic
relaxation. In the bottom layer~1 with $\theta=\alpha=0$, the
in-plane displacement ${\bf{u}}^{(1)}({\bf{R}})$ of an atom with
respect to its initial position ${\bf{R}}$ can be represented by a
Fourier expansion as
\begin{equation}
{\bf{u}}^{(1)}({\bf{R}}) = %
\sum_j {\bf{\tilde{u}}}^{(1)}({\bf{G}}^{(s)}_j)
\sin({\bf{G}_j}^{(s)}{\cdot}{\bf{R}}) \,. %
\label{eqS3}
\end{equation}
The sum extends over all vectors ${\bf{G}_j^{(s)}}$ of the
reciprocal Moir\'{e} superlattice of the BLG, which is spanned by
${\bf{b}}_1^{(s)}$ and ${\bf{b}}_2^{(s)}$.

In the twisted and sheared top layer~2, the in-plane displacement
${\bf{u}}^{(2)}({\bf{R}})$ of an atom with respect to its position
${\bf{R}}$ in the unrelaxed, unsheared, but twisted layer can be
expressed by
\begin{eqnarray}
{\bf{u}}^{(2)}({\bf{R}}) &=& %
  \sum_j {\bf{\tilde{u}}}^{(2)}({\bf{G}}^{(s)}_j)
  \sin({\bf{G}}^{(s)}_j{\cdot}S_\alpha {\bf{R}}) \nonumber \\%
&+& S_\alpha {\bf{R}} - {\bf{R}} \,. %
\label{eqS4}
\end{eqnarray}
Here, the direction angle $\beta$ of the shear transformation,
defined as $S_\alpha$ in Eq.~(\ref{eqS1}), has been rotated by the
twist angle $\theta$. We note that atomic displacements described
by Eqs.~(\ref{eqS3}) and (\ref{eqS4}) maintain the shape and size
of the Moir\'{e} supercells. In BLG subject to twist angles near
${\theta_m}=1.08^\circ$, the summation in Eqs.~(\ref{eqS3}) and
(\ref{eqS4}) requires only three shortest vectors of the
reciprocal Moir\'{e} superlattice
${\bf{G}_1^{(s)}}={\bf{b}}_1^{(s)}$, %
${\bf{G}_2^{(s)}}={\bf{b}}_2^{(s)}$, and %
${\bf{G}_3^{(s)}}=-{\bf{b}}_1^{(s)}-{\bf{b}}_2^{(s)}$. %
In BLG subject to smaller twist angles $\theta$ around
$0.4^\circ$, we include three additional vectors
${\bf{G}_4^{(s)}}=2{\bf{b}}_1^{(s)}$, %
${\bf{G}_5^{(s)}}=2{\bf{b}}_2^{(s)}$, and %
${\bf{G}_6^{(s)}}=-2{\bf{b}}_1^{(s)}-2{\bf{b}}_2^{(s)}$ %
for convergence.

Expressions for the in-plane atomic displacements
${\bf{u}}^{(n)}({\bf{R}})$ in layers $n=1,2$ allow us to determine
the shift vector $\bm{\delta}({\bf{R}})$ at the position
${\bf{R}}$ by
\begin{equation}
\bm{\delta}({\bf{R}}) = %
\left[ {\bf{u}}^{(2)}(T_\theta {\bf{R}}) - %
       {\bf{u}}^{(1)}({\bf{R}}) \right] %
+ \left[ T_\theta {\bf{R}} - {\bf{R}} \right] \,. %
\label{eqS5}
\end{equation}
The periodicity of ${\bf{u}}^{(1)}({\bf{R}})$ and
${\bf{u}}^{(2)}({\bf{R}})$ of the Moir\'{e} superlattice,
described by the Fourier sum in Eqs.~(\ref{eqS3}) and
(\ref{eqS4}), describes the same periodic behavior of
$\bm{\delta}({\bf{R}})$.

Similarly, the periodicity in the interlayer separation
$d_0({\bf{R}})$ can be evaluated by the Fourier sum
\begin{equation}
d_0({\bf{R}})= %
<d_0>+{\Delta}{d_0}{\sum_{j=1}^3}\cos({\bf{G}_j^{(s)}}{\cdot}{\bf{R}}) \,.%
\label{eqS6}
\end{equation}
Here, $<d_0>$ is the average interlayer distance and ${\Delta}d_0$
describes the modulation of $d_0$. Since $d_0$ varies smoothly
across the Moir\'{e} supercell, an adequate description of
$d_0({\bf{R}})$ can be obtained using only three shortest vectors
${\bf{G}_j^{(s)}}$ of the reciprocal Moir\'{e} superlattice
defined above for $j=1-3$. Having specified the position
dependence of the interlayer distance $d_0({\bf{R}})$, the
out-of-plane displacement $h^{(n)}$ of each individual layer $n$,
used to evaluate $E_{el}$ using Eq.~(\ref{eq1}), is given by %
\begin{eqnarray}
h^{(1)}({\bf{R}})&=&\frac{1}{2}
        \left(-d_0({\bf{R}})+3{\Delta}d_0+<d_0>\right)~~{\textrm{and}} \\ %
\label{eqS7} h^{(2)}({\bf{R}})&=&\frac{1}{2}
        \left(+d_0({\bf{R}})-3{\Delta}d_0-<d_0>\right)\,.
\label{eqS8}
\end{eqnarray}

With the Fourier expansions of the in-layer displacement
${\bf{u}}^{(n)}({\bf{R}})$ and the interlayer distance
$d_0({\bf{R}})$, we can determine the elastic energy $E_{el}$ of
STBLG using Eq.~(\ref{eq1}).

To determine the interlayer interaction energy $E_{int}$ specified
in Eq.~(\ref{eq2}), we have to locate a proper expression for the
interlayer interaction potential $V({\bf{R}})$.
Since $V({\bf{R}})$ is periodic in STBLG, we can also express it
as a Fourier expansion~\cite{Zhou2015}
\begin{equation}
V({\bf{R}}) = %
\sum_j \tilde{V}({\bf{G}}^{(s)}_j)
\cos({\bf{G}_j}^{(s)}{\cdot}{\bf{R}}) \,. %
\label{eqS9}
\end{equation}
The sum extends over all vectors ${\bf{G}_j^{(s)}}$ of the
reciprocal Moir\'{e} superlattice of the BLG. In reality,
$V({\bf{R}})=V({\bm{\delta}}({\bf{R}}),d_0({\bf{R}}))$ depends not
on the global position ${\bf{R}}$, but rather the quantities
${\bm{\delta}}$ and $d_0$, which show the same periodicity, as
expressed in Eqs.~(\ref{eqS5}) and (\ref{eqS6}). Using
straight-forward algebra, we arrive at the expression
\begin{eqnarray}
V({\bm{\delta}},d_0) &=& V_{AA}(d_0)-6{\Delta}V(d_0) \nonumber \\
                     &+& 2{\Delta}V(d_0) \sum_{j=1}^{3}
                     \cos({\bf{G}}_j{\cdot}{\bm{\delta}})\,, %
\label{eqS10}
\end{eqnarray}
%
where the sum is limited to the smallest three vectors
${\bf{G}}_1=\bf{b}_1$, %
${\bf{G}}_2=\bf{b}_2$, and %
${\bf{G}}_3=-\bf{b}_1-\bf{b}_2$, since $V$ varies smoothly with
${\bm{\delta}}$.

\begin{figure}[b]
\includegraphics[width=0.75\columnwidth]{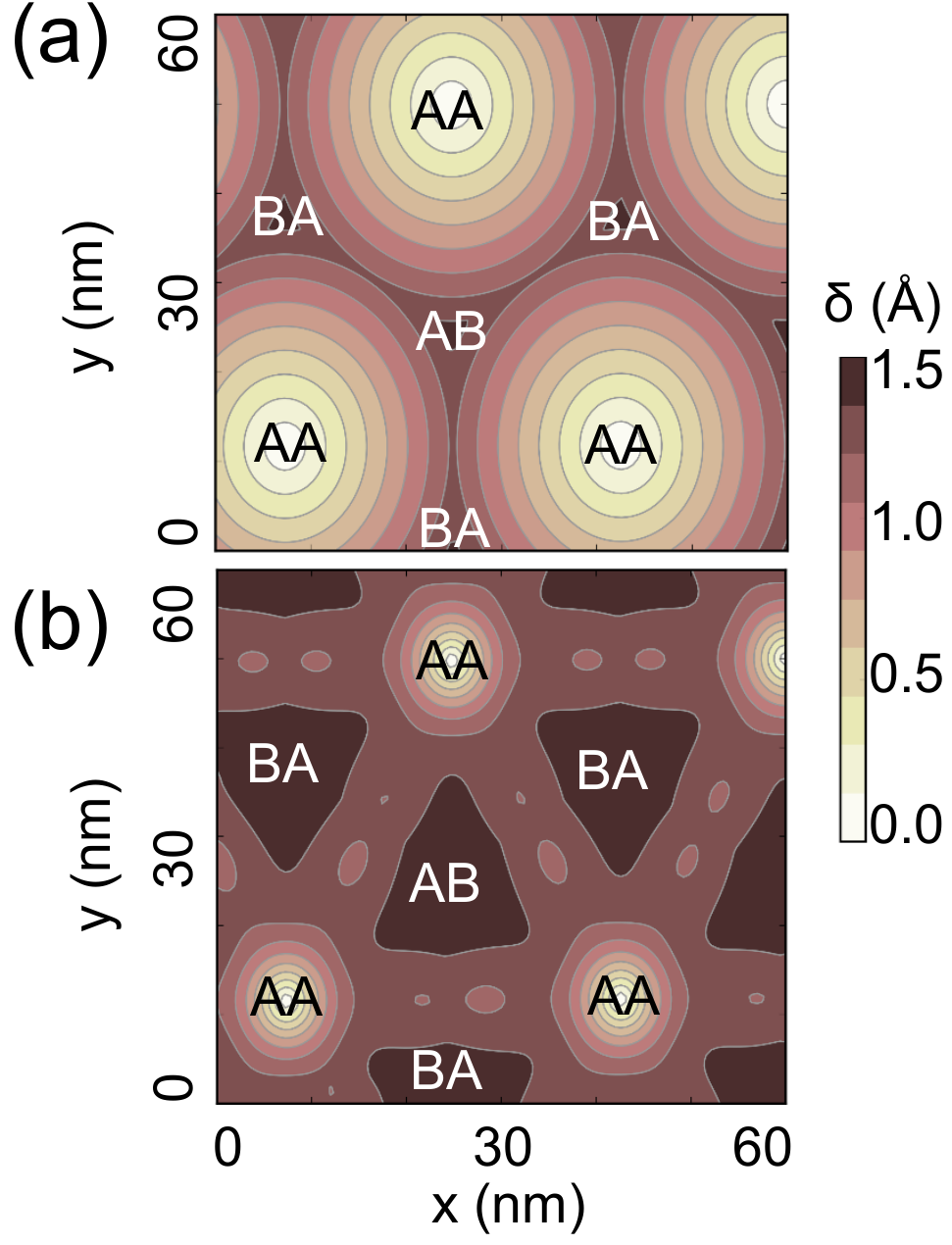}
\caption{Contour plots of the local shift vector length
${\delta}=|\bm{\delta}|$ in the BLG structure, which had been
twisted by $\theta=0.4^\circ$ and sheared by
$\alpha=0.08^\circ$ along the $\beta=0^\circ$ direction, %
(a) in absence and %
(b) in presence of lattice relaxation. %
\label{fig4} }
\end{figure}
\begin{figure}[t]
\includegraphics[width=1.0\columnwidth]{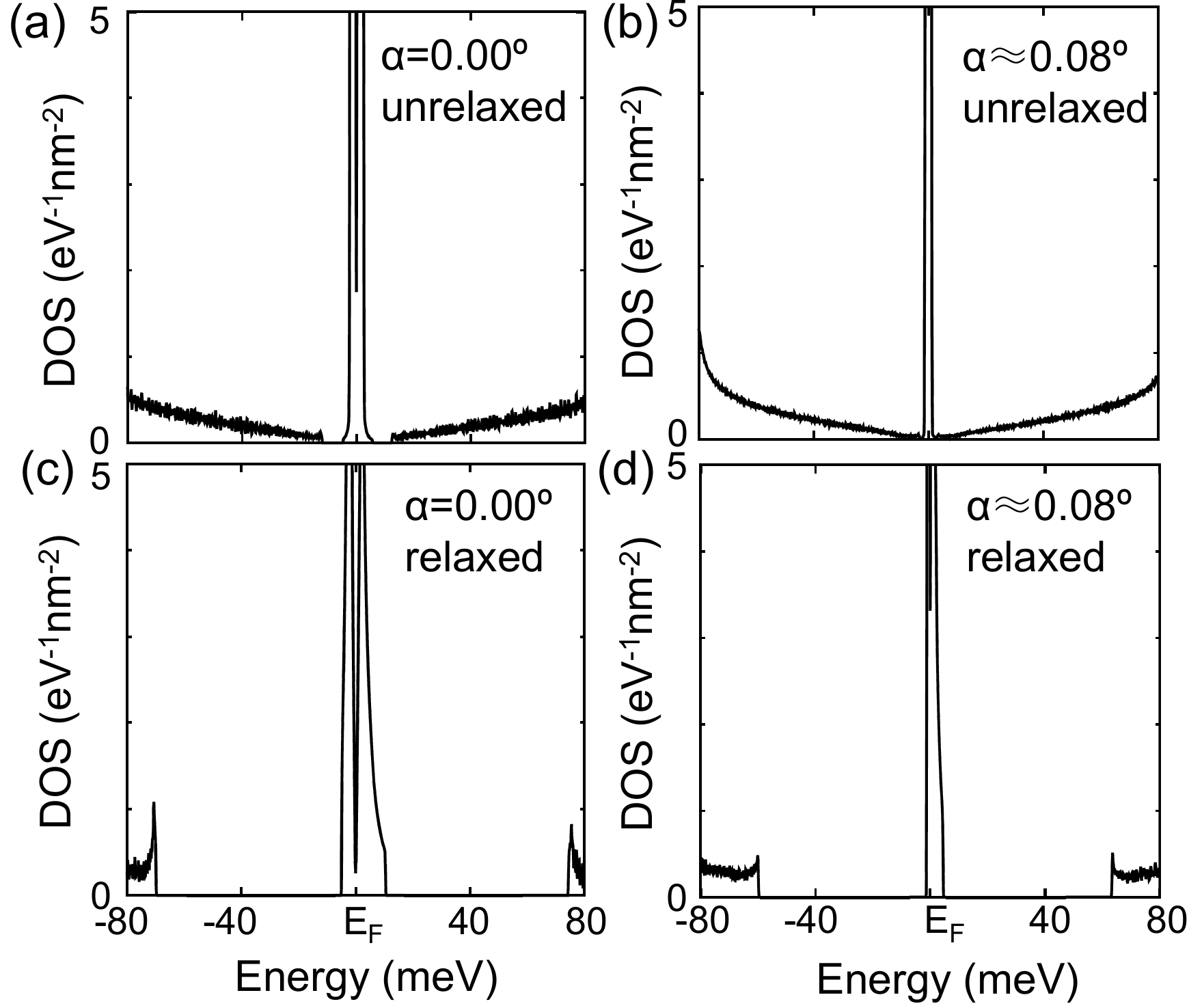}
\caption{Electronic density of states (DOS) in the BLG structure
subject to the magic angle twist $\theta_m=1.08^\circ$. Results
for the unsheared structure are shown in panels (a) and (c), and
those for the top layer sheared by $\alpha=0.08^\circ$ along the
$\beta=0^\circ$ direction in (b) and (d). Results for the
unrelaxed structure in (a) and (b) are compared to those for the
relaxed structure in (c) and (d). The energy scale is extended in
comparison to Fig.~\ref{fig3}. %
\label{fig5} }
\end{figure}

We have used the expression %
${\Delta}V(d_0)=(1/9)[V^{AA}(d_0)-V^{AB}(d_0)]$ for the periodic
modulation of $V$ in Eq.~(\ref{eqS10}), where $V^{AA}(d_0)$ and
$V^{AB}(d_0)$ are the total energies of AA- and AB-stacked BLG per
area, which both depend on the interlayer distance $d_0$.
Furthermore, we have set $V^{AA}(d_0^{AA})=0$ as a reference
value. Using our DFT results, we have fitted $V_{AA}(d_0)$ and
$V_{AB}(d_0)$ by third-order polynomials. The expressions in units
of eV/{\AA}$^2$ are %
${V^{AA}}(d_0)=[0.113{(d_0-d_0^{AA})^2}-%
0.340{(d_0-d_0^{AA})^3}]/\Omega_0$ and %
${V_{AB}}(d_0)=[-0.020+0.174{(d_0-d_0^{AB})^2}-%
0.224{(d_0-d_0^{AB})^3}]/\Omega_0$ %
when using $d_0$ in {\AA} units and the value
$\Omega_0=5.24$~{\AA}$^2$ for the area of the graphene unit cell.

With expressions for $E_{el}$ and $E_{int}$ in place, we can
evaluate the total energy $E_{tot}=E_{el}+E_{int}$ for any BLG
geometry. For given global shear characterized by $\alpha$ and
$\beta$ and given global twist given by $\theta$, we can determine
the atomic relaxations by globally minimizing the total energy
with respect to $\bf{\tilde{u}}^{(n)}({\bf{G}}^{(s)})$,
${\Delta}d_0$ and $<d_0>$.


\subsection{Local relaxation in sheared and twisted bilayer graphene}

The pattern of the local shift vectors $\bf\delta$ depends
primarily on the twist angle $\theta$, which determines the domain
size. The pattern in the bilayer graphene (BLG) lattice twisted by
a very small angle $\theta=0.4^\circ$ is presented in
Fig.~\ref{fig4}. The domains of the Moir\'{e} superlattice are
significantly larger than in BLG with the magic twist angle
${\theta}_m=1.08^\circ$ discussed earlier.
The $\delta$ pattern in the larger domains is changed
substantially by atomic relaxation. In particular, the domains of
AB or BA stacking acquire a distinct triangular character that has
been observed in dark-field TEM images\cite{McEuenBLG13,yoo18}.

\begin{figure}[h]
\includegraphics[width=0.7\columnwidth]{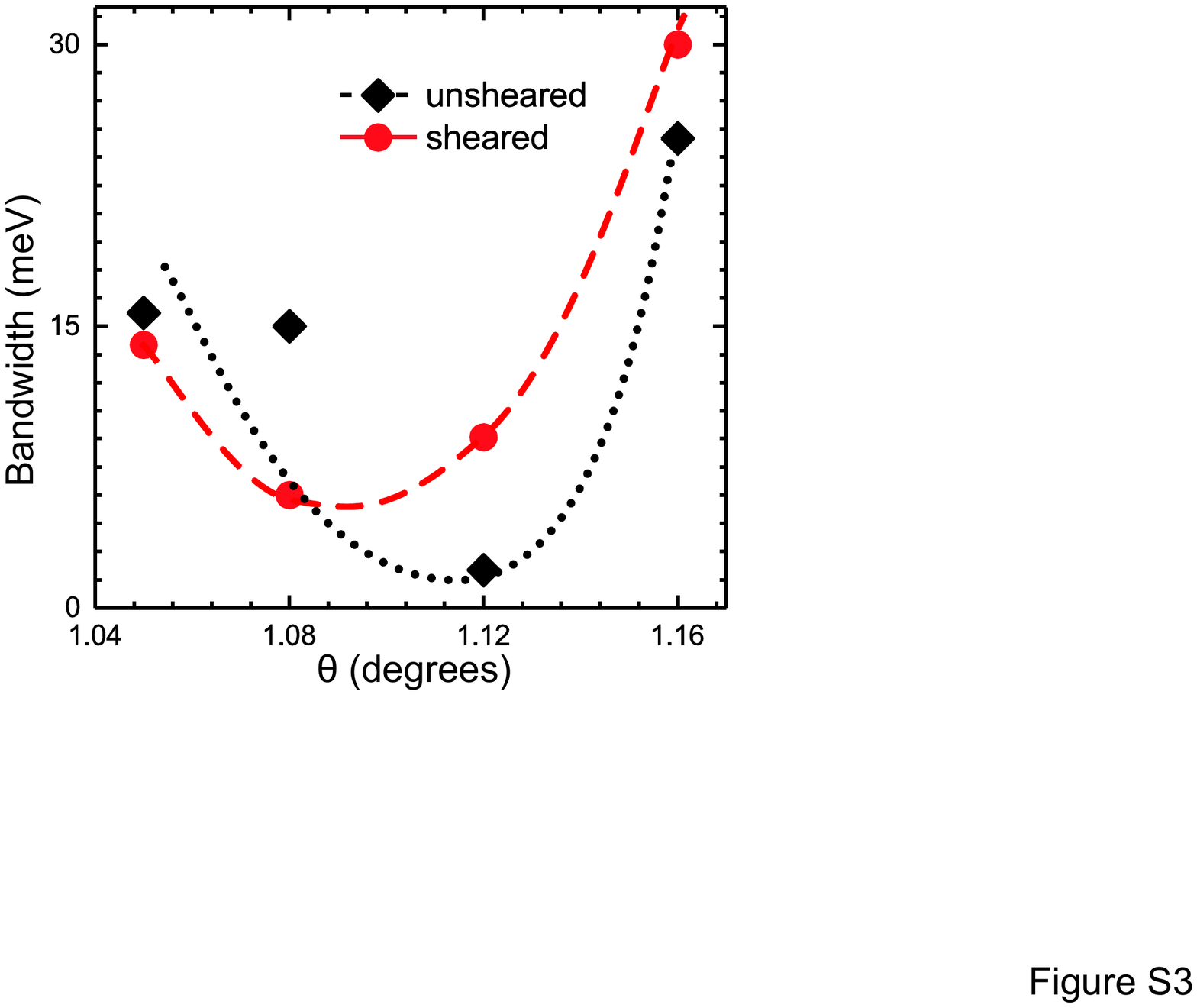}
\caption{The width of the 'flat band' in a relaxed twisted bilayer
graphene structure as a function of the twist angle $\theta$. Data
points for commensurate, relaxed structures in the vicinity of the
observed value ${\theta_m}=1.08^\circ$ are presented by the black
diamonds ($\blacklozenge$) for the unsheared lattice and by the
red circles ($\bullet$) for the structure with the top layer
sheared by ${\alpha}{\approx}0.08^\circ$ along the
${\beta}=0^\circ$
direction. The dotted and dashed lines are guides to the eye. %
\label{fig6} }
\end{figure}

\subsection{Electronic structure changes in sheared and twisted
             bilayer graphene}

The electronic density of states (DOS) in sheared and twisted BLG
is shown in Fig.~\ref{fig5} in a larger energy window around $E_F$
than in Fig.~\ref{fig3}.
We note that both atomic relaxation and global shear modify the
spectrum of the unsheared and unrelaxed lattice, shown in
Fig.~\ref{fig5}(a), significantly. Band gaps above and below the
`flat band' region are significant only in the relaxed structure.
Only the DOS of the sheared and relaxed structure in
Fig.~\ref{fig5}(d) displays a combination of a narrow flat band
and band gaps above and below.

\subsection{Value of the magic angle in unsheared and sheared
             twisted bilayer graphene}

Since shear plays an important role in the relaxed twisted bilayer
graphene, it is expected to affect the magic angle $\theta_m$ as
well. In Fig.~\ref{fig6} we show the width of the narrow band
around $E_F$ as a function of the twist angle $\theta$ for both an
unsheared BLG and for BLG with its top layer sheared by
${\alpha}{\approx}0.08^\circ$ along the ${\beta}=0^\circ$
direction. Only few data points are shown, since only few
commensurate structures with reasonably small unit cells exist in
the narrow twist angle region shown. We notice that the (first)
magic angle, associated with the narrowest bandwidth, occurs at
$1.12^\circ$ in the unsheared and at $1.08^\circ$ in the globally
sheared BLG. The latter value in the sheared lattice agrees well
with the observed value~\cite{{Cao2018},{cao2018unconventional}}
$\theta_m({\rm{expt}}){\approx}1.08^\circ$.


\label{Acknowledgments}
\begin{acknowledgments}
D.T. and D.L. acknowledge financial support by the NSF/AFOSR EFRI
2-DARE grant number EFMA-1433459. X.L. acknowledges support by the
China Scholarship Council. Computational resources have been
provided by the Michigan State University High Performance
Computing Center.
\end{acknowledgments}


%

\end{document}